\def\aj{{AJ}}

\def\apj{{ApJ}}
\def\apjs{{ApJS}}

\def\mnras{{MNRAS}}

\def\pasp{{PASP}}
\def\kms{{km\,s$^{-1}$}}
\def\kmsmpc{{km\,s$^{-1}$\,Mpc$^{-1}$}}
\def\Ho{{$H_{\rm 0}$}}

\def\lmcdm{{$\mu_{\rm LMC}$}}
\def\araa{{ARA\&A}}
\def\aap{{A\&A}}

\documentclass[cup5b]{caps}
\usepackage{graphicx}
\usepackage{amssymb}
\usepackage{ociwsymp2}
\HeadText{J. B. Jensen, J. L. Tonry, and J. P. Blakeslee}

\begin{document}

\pagenumbering{arabic}

\author[]{J. B. JENSEN$^{1}$, J. L. TONRY$^{2}$, and J. P. BLAKESLEE$^{3}$\\
(1) Gemini Observatory \\
(2) University of Hawaii Institute for Astronomy \\
(3) John Hopkins University}

\chapter{The Extragalactic Distance Scale}

\begin{abstract}
Significant progress has been made during the last 10 years
toward resolving the debate over the expansion rate of the Universe.
The current value of the Hubble parameter, \Ho, is now arguably known
with an accuracy of 10\%, largely due to the tremendous increase in
the number of galaxies in which Cepheid variable stars have been
discovered.  
Increasingly accurate secondary distance indicators, many calibrated
using Cepheids, now provide largely concordant measurements of \Ho\ well 
out into the Hubble flow, and deviations from the smooth Hubble flow allow 
us to better measure the dynamical structure of the local Universe.  
The change in the Hubble parameter with redshift provided the first 
direct evidence for acceleration and ``dark energy'' in the Universe.  
Extragalactic distance measurements are central to determining the
size, age, composition, and fate of the Universe.
We discuss remaining systematic uncertainties, particularly related
to the Cepheid calibration, and identify where improvements are
likely to be made in the next few years.
\end{abstract}

\section{Introduction}

The measurement of distances to the ``nebulae'' early in the twentieth
century revolutionized our
understanding of the scale of the Universe and provided the first
evidence for universal expansion (an overview of the history of
cosmology can be found in Longair (this volume).
Distance measurements have played a profound role in unraveling the 
nature of the Universe and the objects in it ever since.
Without knowing how far away objects are, we would not be able to learn
very much about their sizes, energy sources, or masses.  
On a universal scale, extragalactic distance measurements lie at the 
heart of our understanding of the size, age, composition, evolution, 
and future of the Universe.
The value of the Hubble parameter today, \Ho, sets the scale of the 
Universe in space and time, and
measuring \Ho\ depends heavily on accurate extragalactic distance
measurements out to hundreds of Mpc.
Accurate distance measurements are also needed to map the deviations
from the smooth Hubble expansion, or peculiar velocities,
which presumably arise gravitationally due to the distribution of 
mass in the local Universe.

During the last decade several important steps have been taken toward
resolving the factor of ${\sim}2$ uncertainty in the scale of the 
Universe that overshadowed observational cosmology for 30 years.
Many of the improvements are the direct results of a better 
calibration and extension of the Cepheid variable star distance scale.
While significant systematic uncertainties remain, 
Hubble constant measurements made using a wide variety of 
distance measurement techniques are now converging on
values between 60 and 85 \kmsmpc; very few measurements lie outside
this range, with the majority falling between 70 and 75 \kmsmpc.
Several recent advances made the improved Cepheid calibration possible.
First, the {\it Hipparcos} satellite made parallax distance measurements 
accurate to 10\% to Cepheid variable stars in the solar neighborhood.  
{\it Hipparcos} parallax measurements helped pin down the 
zeropoint brightness of the Cepheid variable stars, but not enough
Cepheids could be observed to properly determine the slope of the
period-luminosity (PL) relation.
Second, the {\it Hubble Space Telescope (HST)}\ provided the spatial resolution 
and sensitivity to detect Cepheids in galaxies as far away as 25 Mpc,
allowing for the first time a calibration of a number 
of secondary extragalactic distance indicators using Cepheids.
Finally, the OGLE microlensing experiment turned up thousands of 
Cepheids in the Large Magellanic Cloud (LMC)
that were used to accurately determine the PL relation.
The improved Cepheid calibration of secondary distance indicators
has largely yielded concordance on the scale of the Universe, 
and the uncertainty in \Ho\ is now arguably close to 10\%.

This summary is not intended to be an inclusive review of all the distance
measurement techniques and their relative strengths and weaknesses.
Instead, we highlight some recent measurements
and identify the most significant remaining systematic uncertainties.
Cepheids are emphasized, as they are currently used to calibrate most
secondary distance indicators.  
We close by summarizing how planned future projects will improve
our knowledge of the expansion rate and eventual fate of the 
Universe.

\section{The Cepheid Calibration}

The Hubble Constant Key Project (KP) team set out to determine the 
Hubble constant to 10\% or better by reliably measuring Cepheid 
distances to galaxies reaching distances of ${\sim}25$~Mpc
(Mould et al. 2000; Freedman et al. 2001).  
The KP sample galaxies included field and cluster spirals, including
several in the nearby Virgo, Fornax, and Leo I clusters.  
The KP team performed $V$ and $I$ band photometry using two 
independent reduction packages and analysis techniques to
understand and control systematic errors as much as possible.
Near-IR measurements with NICMOS were used to check the validity
of the reddening law adopted by the KP team (Macri et al. 2001).
All the KP results have been presented using
a distance modulus to the LMC of 18.50 mag.
To keep the KP results on a common footing, the KP measurements 
were all reported using the PL relation
determined by Freedman \& Madore (1990), which was derived from
a relatively limited set of LMC Cepheids.
No adjustment to the PL relation was made to the baseline KP
measurements for differences in 
metallicity between Cepheids (Ferrarese et al. 2000b).

The KP team was not alone in taking advantage of {\it HST}'s
spatial resolution and sensitivity
to find Cepheids in relatively distant galaxies.
Cepheids have also been measured in supernova (SN) host galaxies by 
A. Sandage, A. Saha, and collaborators.
Their observations targeted galaxies in which Type Ia SNe have 
occurred for the purpose of calibrating SNe as a standard candle.  
The Sandage and Saha team made use of similar
data reduction techniques as the KP team and the same LMC distance.
An additional Cepheid measurement in the Leo I galaxy NGC 3368 was 
made by Tanvir, Ferguson, \& Shanks (1999); NGC 3368 later hosted SN 1998bu.

Data for the entire combined sample of 31 Cepheid galaxies
from both teams was presented
by Ferrarese et al. (2000b) and Freedman et al. (2001) 
to facilitate comparison and calibration of
secondary distance indicators on a common Cepheid foundation
(Ferrarese et al. 2000a,b; Gibson et al. 2000;
Kelson et al. 2000; Sakai et al. 2000).  
Mould et al. (2000) combined these results and found 
\Ho$=71\,{\pm}\,6$ \kmsmpc.
The results of the KP calibration of the
SN, Tully-Fisher (TF), fundamental plane (FP), and surface brightness 
fluctuation (SBF) distance indicators are included in the subsequent 
sections.

The gravitational lensing data of the OGLE experiment increased the
number of good Cepheid measurements in the LMC by more than an 
order of magnitude (${\sim}650$ fundamental-mode Cepheids; 
Udalski et al. 1999a,b).
Freedman et al. (2001) applied the new Cepheid PL relation to
the combined Cepheid database.
The same LMC distance modulus of \lmcdm$=18.50$ mag used in the earlier
KP papers was maintained by Freedman et al.
They also argued for a modest metallicity correction of 
$-0.2\,{\pm}\,0.2$ mag dex$^{-1}$ in (O/H) (Kennicutt et al. 1998).
It is interesting to note that the new greatly improved PL relation 
has a somewhat different slope in the $I$ band, resulting in a 
distance-dependent offset.  Only the brightest, longest-period 
Cepheids can be detected in the most distant galaxies, so the 
revised slope will have the largest effect in the most distant
galaxies.
Adopting the new PL relation reduces the distance moduli of the
most distant galaxies by up to ${\sim}0.2$ mag.
The metallicity correction counteracts
the shorter distances to some extent, and the change in the
resulting Hubble constant when adopting both the metallicity correction
and the new PL relation is small ($72\,{\pm}\,8$ \kmsmpc).  
If the new PL relation
is adopted without the metallicity correction, the Hubble constant
would increase by a few percent (depending on which Cepheid galaxies
are used to calibrate a particular secondary distance indicator).
An recent independent survey of LMC Cepheids has confirmed the slope of
the OGLE PL relation (Sebo et al. 2002).  
The new results, derived from Cepheids with periods comparable to those
of the more-distant KP galaxies, agrees with the Udalski et al. (1999a,b) 
OGLE PL zeropoint to 0.04 mag (or 2\% in \Ho), well within the
uncertainties of the two measurements. 

In the following sections, all Cepheid-based distance measurements will
be compared to the Ferrarese et al. (2000b) scale, using the original
KP zeropoint and no metallicity correction, or to the Freedman et al.
(2001) compilation, which uses the OGLE (``new'') PL relation and 
metallicity correction of ${-}0.2$ mag dex$^{-1}$.
In all cases, the LMC distance adopted is 18.50 mag.
The metallicity-corrected Freedman et al. (2001) and 
uncorrected Ferrarese et al. (2000b) KP compilations
are not strictly comparable; the metallicity difference between
Galactic and LMC Cepheids would result in a ${\sim}0.08$ mag 
difference in the distance modulus to the LMC.

\section{Secondary Distance Indicators and the Hubble Constant}

Most secondary indicators derive their zeropoint calibration
from Cepheids.  We focus here on those techniques that have been
calibrated using the common foundation of the KP Cepheid 
measurements.  A few new results that are independent of the
Cepheid calibration are presented as well (Table~\ref{table1}).

\subsection{Type Ia Supernovae}

The brightness of exploding white dwarf SNe can be calibrated 
using a single parameter (Phillips 1993; Hamuy et al. 1995; 
Riess, Press, \& Kirshner 1996).  
After correcting their luminosities for decline rates, Type Ia 
SNe are a 
very good standard candle with a variance of about 10\%.
Both the KP and Sandage and Saha teams have calibrated Ia SNe using 
{\it HST}\ Cepheid measurements.
While many of the Cepheids observations are identical, the two teams
make numerous different choices regarding the detailed analyses.
They also make use of different historical SNe to compute 
the value of \Ho.  
The Sandage and Saha team consistently get larger distances and smaller
values of the Hubble constant than the KP team does.  The differences
are discussed in detail by Parodi et al. (2000) and Gibson et al.
(2000).  Parodi et al. find \Ho$=58\,{\pm}\,6$ \kmsmpc, while Gibson et al.
report \Ho$=68\,{\pm}\,2\,{\pm}\,5$ \kmsmpc\ (the first uncertainty
is statistical, the second systematic).  
Both teams used the same LMC distance and
similar reduction software, but included different calibrators
and different analysis techniques.
Much of the disagreement between the two groups
has its origin in the selection and analysis of the individual 
Cepheid variables and which to exclude.  The remaining difference
arises from the choice of which historical SN data to trust and
which SNe to exclude from the fit to the distant Hubble flow.
Hamuy et al. (1996) measured a value of \Ho$=63\,{\pm}\,3\,{\pm}3$
\kmsmpc\ using the 30 Ia SNe of the Calan/Tololo survey and four 
Cepheid calibrators.
Ajhar et al. (2001) found that the optical $I$-band SBF 
distances to galaxies with Type Ia SNe were entirely consistent 
when differences between Cepheid calibrators were taken into account.
They found that SBFs and SNe give identical values of \Ho\ [73 \kmsmpc\
on the original KP system, 75 on the new Freedman et al. (2001) calibration, 
and 64 on the Sandage and Saha calibration].

SNe at high redshift ($z\,{>}\,0.5$), and their departure
from a linear Hubble velocity, have been used to
explore the change in the universal expansion rate with time.
The measurements of two collaborations (the High-z Team, led by 
B. Schmidt, and the Supernova Cosmology Project, led by S.
Perlmutter) have provided the best
evidence to date that the Universe is expanding at an increasing
rate.  The implication of the SN data is that ${\sim}$70\% of 
the energy density in the Universe is in some form of ``dark energy'' 
such as vacuum energy, ``quintessence,'' or something even more bizarre
(Perlmutter et al. 1997, 1998; Schmidt et al. 1998).  
The use of SNe to probe the equation of state of the Universe is the 
topic of other papers in this volume.  In general, the distant SNe do 
not need to be put on an absolute distance scale to study the
change in the Hubble parameter with time.

Kim et al. (1997) used the first few SNe discovered by the Supernova
Cosmology Project to constrain the Hubble constant.  They found that 
\Ho$\,{<}\,82$ \kmsmpc\ in an 
$\Omega_{\Lambda}\,{=}\,0.7, \Omega_{m}\,{=}\,0.3$ 
Universe (as suggested by the distant SN data and cosmic microwave
background measurements of the flatness of the Universe).

\begin{figure}
\centering
\includegraphics[height=85mm,width=115mm]{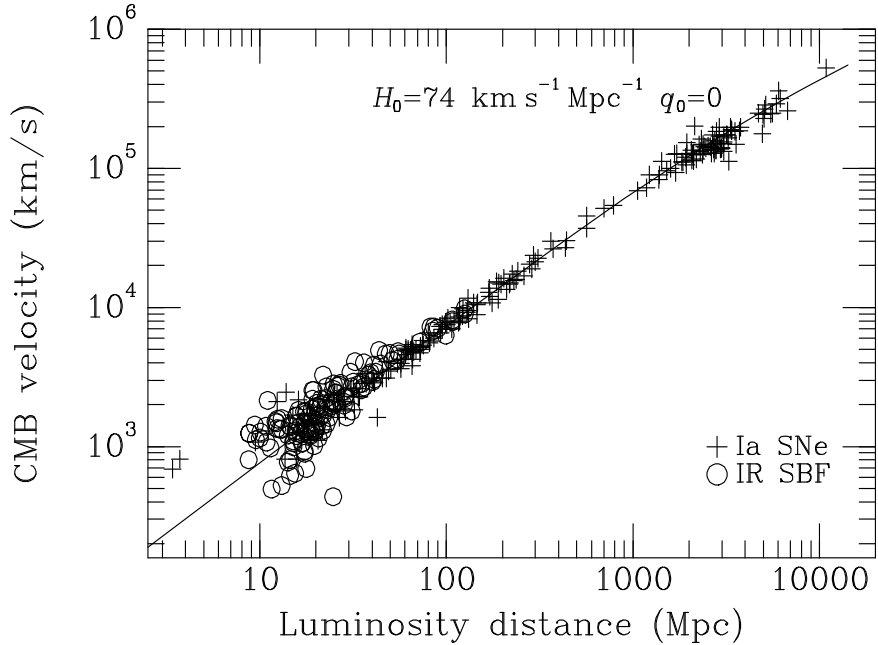}
\caption{Hubble diagram derived from the combined IR SBF and Type Ia
SN data (Jensen et al. 2003; Tonry et al. 2003).  The line indicates
the expansion velocity for an empty Universe.  The fit to the SN
data give a Hubble constant of 74 \kmsmpc.}
\label{hubblediagram}
\end{figure}

Tonry et al. (2003, in preparation) have recently compiled a 
database of all the currently available Type Ia SNe data using a 
common calibration and consistent analysis techniques.
The result is a uniform data set of 209 well-measured SN distances
in units of \kms.  An independent distance to any of the galaxies 
therefore leads immediately to a tie to the Hubble flow out to redshifts
greater than one. 
Six galaxies from the Tonry et al. (2003) database have Cepheid 
distances determined by the KP team (Freedman et al. 2001).
Using the new PL relation and metallicity corrections,
the SN data give a Hubble constant of $74\,{\pm}\,3$ \kmsmpc.  
The exquisite tie to the Hubble flow is shown in 
Figure~\ref{hubblediagram}, along with the best fit Hubble constant.
The line indicates the evolution of \Ho\ for an empty ($q_0\,{=}\,0$)
cosmology, and the deviation from that line at $z\,{\approx}\,0.5$ to 1
is the best evidence for an accelerating ``dark energy'' dominated
Universe.  

The distant SN data more tightly constrain the product 
$H_0t_0$ than \Ho\ alone.
The long, thin error ellipses of the SN data fall along lines
of constant $H_0t_0$ for SNe at $z{\approx}0.5$.
The Supernova Cosmology Project found $H_0t_0\,{=}\,0.93\,{\pm}\,0.06$ 
(Perlmutter et al. 1997), and the High-z team
measured $H_0t_0\,{=}\,0.95\pm0.04$ (Tonry et al. 2003).  
For an age of the Universe of 13 Gyr, this implies a
Hubble constant of ${\sim}70$ \kmsmpc.  In the future, other constraints
on the age of the Universe, combined with highly accurate values of
$H_0t_0$ from high-redshift SNe at different redshifts, 
will give us better constraints on the Hubble parameter and how it has
changed with time.

\subsection{Surface Brightness Fluctuations}

The amplitude of luminosity fluctuations in dynamically hot systems
arises due to statistical fluctuations in the number of stars per
resolution element (Tonry \& Schneider 1988; Blakeslee, Ajhar, \& Tonry 1999).  
SBFs are distance-dependent:  the nearer a galaxy
is, the bumpier it appears.  The brightness of the fluctuations depends 
directly on the properties of the brightest stars in a given population,
making SBFs a stellar standard candle.  Significant SBF surveys have been
completed at $I$, where the effects of age and metallicity are degenerate,
and in the near-IR, where fluctuations are brightest and extinction is
minimized.

The Hubble constant derived from $I$-band 
SBFs, as calibrated by the KP team (Ferrarese et al. 2000a), is
\Ho$\,{=}\,70\,{\pm}\,5\,{\pm}\,6$ \kmsmpc (random and systematic
uncertainties, respectively), 
using the four distant {\it HST}\ measurements of Lauer et al. (1998).
The $I$-band SBF team used the much larger sample of ${\sim}300$ 
galaxies and a slightly different calibration to 
find a somewhat larger Hubble constant of \Ho$=77\,{\pm}\,4\,{\pm}\,7$ 
(Tonry et al. 2000, 2001), using the
original KP calibration of Ferrarese et al. (2000b).
The $I$-band SBF survey team fitted a detailed model of the velocity
field of the local Universe to get their determination of the 
Hubble constant. 
Half the difference between the Ferrarese et al. (2000a) and 
Tonry et al. (2000)
results is due to the velocity field corrections, and the other to
differences in the choice of Cepheid calibration galaxies.
SBFs can be measured in the bulges of a few spiral galaxies with known
Cepheid distances, and the preferred SBF calibration of Tonry et al.
uses only galaxies with distances known from both Cepheids and SBFs.
As with all the secondary indicators calibrated using Cepheids, 
moving to the new PL relation would result in an increase in the
Hubble constant of 3\% including the Freedman et al. (2001) metallicity
correction, or 8\% using the new PL relation alone.

Infrared measurements using NICMOS on the {\it HST}\ have extended SBF 
measurements beyond 100 Mpc, where deviations from the smooth Hubble
flow should be small.  Jensen et al. (2001) measured a Hubble constant
between 72 and 77 \kmsmpc\ using the original KP Cepheid calibration.
When reanalyzed using an updated calibration and the new OGLE PL
relation, they find \Ho$\,{=}\,77$ \kmsmpc\ using the 
Freedman et al. (2001) calibration
(including the metallicity correction; Jensen et al. 2003).

\subsection{Fundamental Plane}

Elliptical galaxies are very homogeneous in their photometric and 
dynamical properties.  By accurately measuring surface brightness,
size, and central velocity dispersion, the position of an elliptical
galaxy on the ``fundamental plane'' (FP) gives an estimate of the distance 
with an accuracy of ${\sim}20$\%.  The FP is an improved version of the
Faber-Jackson and $D_N-\sigma$ relations, which are also
used to determine distances to elliptical galaxies.
Kelson et al. (2000) combined various FP data for the Fornax, 
Virgo, and Leo I clusters, for which Cepheid distances had been
measured.
They applied the Cepheid calibration of the FP relation for the three
nearby clusters to 11 more distant clusters.
The resulting Hubble constant of $82\,{\pm}\,5\,{\pm}\,10$ \kmsmpc\ 
is somewhat higher than the KP estimates using other secondary
distance indicators.
Adopting the metallicity correction of Kennicutt et al. (1998)
would reduce the value of \Ho\ by 6\%.
The relative placement of the spiral Cepheid calibrators and elliptical
FP galaxies within the three nearby clusters is one of the primary 
sources of systematic uncertainty (${\sim}5$\%).

Hudson et al. (2001) have combined a number of FP data sets, making them 
much more homogeneous by cross checking the photometry and velocity
dispersion measurements.  When the Hudson et al. data are
analyzed using the Ferrarese et al. (2000b) 
Cepheid calibration, Blakeslee et al. (2002) find that the Hubble
constant is consistent with SBFs and with other secondary distance
indicators.  They find \Ho$=68\,{\pm}\,3$ \kmsmpc, in excellent agreement
with the KP calibrations of the other secondary distance indicators.  
Using the new PL relation (Freedman et al. 2001), the FP Hubble
constant is $73\,{\pm}\,4\,{\pm}\,11$ \kmsmpc, where SBFs have been 
used to make a direct connection between the FP galaxies and the 
Cepheid calibrators.

\subsection{Tully-Fisher}

Like elliptical galaxies, the photometric and kinematic properties of 
spiral galaxies are closely related.
The rotation velocities and brightnesses of spiral galaxies 
can be measured, making it possible to estimate the distance to a galaxy.
Tully-Fisher (TF) distance measurements are among the most widely used,
although the accuracy of an individual measurement is generally
taken to be about 20\%.
Sakai et al. (2000) used the compiled TF data for 21 calibrators with
Cepheid distances and applied the results to a large data set of
23 clusters within 10,000 \kms\ (Giovanelli et al. 1997).  
Sakai et al. found that the TF Hubble constant
is \Ho$=71\,{\pm}\,4\,{\pm}\,7$ \kmsmpc,
in very good agreement with the other techniques already mentioned.

\subsection{Type II Supernovae}

The expanding-photospheres method of determining distances to Type II
SNe can be calibrated using a zeropoint based on Cepheid distances 
(although the expanding-photospheres method is a primary distance indicator 
that does not
require a Cepheid calibration, as described in \S~\ref{bypass}).
The KP team (Freedman et al. 2001) applied the new Cepheid calibration 
to the SN measurements of Schmidt et al. (1994) 
and found \Ho$\,{=}\,72\,{\pm}\,9\,{\pm}\,7$ \kmsmpc, 
in close agreement with the value 
of \Ho$\,{=}\,73\,{\pm}\,6\,{\pm}\,7$ reported by Schmidt et al.
A second way of using Type II SNe as a distance estimator has been
developed by Hamuy (2001).  The expansion velocity for a particular
type of ``plateau'' SN is correlated with its luminosity.
The average of four SNe give a Hubble constant of $75\,{\pm}\,7$ \kmsmpc\
using a Cepheid calibration comparable to the Freedman et al. zeropoint
(M. Hamuy, private communication).  The best-measured and only modern
SN of the four (SN 1999em) gives \Ho$\,{=}\,66\,{\pm}\,12$ \kmsmpc.  

\subsection{Other Distance Indicators Calibrated Using Cepheids}

The KP team also presented Cepheid calibrations of several other
distance indicators, including the globular cluster luminosity 
function (GCLF), the planetary nebula luminosity function (PNLF), 
and the tip of the red giant branch (TRGB). 
Ferrarese et al. (2000a,b) compiled results for these techniques
and compared them to SBFs.

The GCLF technique has been recently shown to be as good a distance
indicator as
other secondary techniques when appropriate corrections are
made for completeness, background sources, and luminosity function
width (Kundu \& Whitmore 2001; Okon \& Harris 2002).  
To measure reliable distances, globular clusters fainter than 
the GCLF peak must be detected.  
Some earlier measurements did not go deep enough to
reach the peak luminosity, and the results were less reliable and 
possibly biased toward smaller distances.  By measuring GCLF distances 
relative to the Virgo cluster, and adopting a distance
to Virgo of 16 Mpc as the calibration (which is independent of, but
consistent with, the Cepheid calibration), the resulting Hubble constant
is near 70 \kmsmpc\ (Okon \& Harris 2002).  The Cepheid distance to 
Virgo from Ferrarese et al. (2000a) is 16.1 Mpc; thus, the agreement
between the KP calibration of the GCLF technique and the newer 
measurements is very good.

Ciardullo et al. (2002) recently made a detailed comparison of the 
PNLF technique to the Cepheid distance scale.  The current PNLF
technique makes a correction for metallicity of the host galaxy.
Using a distance of 710 kpc to M31 to determine the zeropoint 
for the PNLF method,
they found that the Cepheid and PNLF distances are consistent within
the statistical uncertainties of the two methods.
The agreement between Cepheids and PNLF in the particular case of
NGC~4258 also leads to a ${\sim}$1-$\sigma$ disagreement with
the geometrical maser distance.

In Table~\ref{table1} we summarize a few recent measurements of \Ho\
and the Cepheid calibration used (when appropriate).
Most of the measurements are calibrated directly using the original
KP zeropoint or the new OGLE PL relation (``New PL''). 
Many also include metallicity corrections (indicated by ``+Z'').
References and additional calibration details are included.
Willick \& Batra (2001) provide an independent calibration of the
Cepheid distance scale using the new OGLE PL relation.

% Table 1:  Ho summary table
\begin{table}
\caption{A Few Recently Published Hubble Constant Measurements}
\scriptsize
%\tiny
\begin{tabular}{llll}
\hline \hline
{Technique} & {\Ho} (\kmsmpc)& {Cepheid calibration} & {Reference} \\
\hline
Key Project summary     & $72\pm8$      & New PL+Z & Freedman et al. 2001\\
Cepheids+IRAS flows     & $85\pm5$      & New PL& Willick \& Batra 2001\\
Type Ia Supernovae      & $59\pm6$      & Sandage team & Parodi et al. 2001\\
                        & $59\pm6$      & Sandage team & Saha et al. 2001\\
                        & $71\pm2\pm6$  & New PL+Z & Freedman et al. 2001\\
                        & $74\pm3$      & New PL+Z & Tonry et al. 2003, in prep.\\
                        & $73\pm2\pm7$  & New PL+Z & Gibson \& Stetson 2001\\
$I$-band SBFs           & $77\pm4\pm7$  & Orig. KP & Tonry et al. 2000\\
                        & $70\pm5\pm6$  & New PL+Z & Freedman et al. 2001\\
                        & $75$          & New PL+Z & Ajhar et al. 2001\\
$H$-band SBFs           & $72\pm2\pm6$  & Orig. KP+$I$-SBF & Jensen et al. 2001 \\
                        & $77\pm3\pm6$  & New PL+Z & Jensen et al. 2003, in prep.\\
$K$-band SBFs           & $71\pm8$      & Orig. KP+$I$-SBF & Liu \& Graham 2001\\
Tully-Fisher            & $71\pm3\pm7$  & New PL+Z & Freedman et al. 2001\\
Fundamental Plane       & $82\pm6\pm9$  & New PL+Z & Freedman et al. 2001\\
                        & $73\pm4\pm11$ & New PL+Z & Blakeslee et al. 2002\\
Type II Supernovae      & $72\pm9\pm7$  & New PL+Z & Freedman et al. 2001\\
                        & $75\pm7$      & New PL+Z & M. Hamuy, private comm.\\
Globular Custer LF      & ${\sim}70$    &similar to Orig. KP & Okon \& Harris 2002\\
Sunyaev-Zel'dovich       & $60\pm3\pm30$\% & ... & Carlstrom et al. 2002\\
Gravitational lenses    & 61 to 65      & ... & Fassnacht et al. 2002\\
                        & $59\pm12\pm3$ & ... & Treu \& Koopmans 2002\\
Type Ia SNe (theory)    & $67\pm9$    & ... & H\"oflich \& Khokhlov, 1996\\
Type II SNe (theory)    & $67\pm9$    & ... & Hamuy 2001\\
\hline \hline

\end{tabular}
\label{table1}
\end{table}

\section{Systematic Uncertainties in the Cepheid Calibration}

The KP and SN calibration teams have 
provided a large and uniform data set
of consistently calibrated Cepheid distances.  There are, however,
several systematic uncertainties that prevent achieving an accuracy 
much better than 10\% in distance.  
The primary systematic uncertainties are common to all the Cepheid
measurements, and are all similar in magnitude 
(Freedman et al. 2001).
A concerted effort to improve the accuracy in several areas is
therefore needed to significantly reduce the uncertainty in \Ho\
using Cepheid-calibrated secondary distance indicators.

\subsection{The Distance to the Large Magellanic Cloud}

The distance to the LMC is a fundamental rung in the distance
ladder.  The LMC is large enough and distant enough to contain a
wide assortment of stellar types at nearly the same
distance, yet close enough for individual stars to be easily
resolved.  The LMC contains stars and globular clusters spanning a wide
range in age.  It also hosted the Type II SN 1987A, the best-studied SN
ever.  The LMC is crucial for the Cepheid
calibration because there are not enough Galactic Cepheids with
independently determined distances to pin down the zeropoint,
the PL relation, and metallicity dependence simultaneously.
Only in the LMC do we have a sample of thousands of Cepheids
at a common distance.

The fact that the distance to the LMC is not yet
well determined is a significant and persistent problem
(Walker 1999;  Paczy\'nski 2001; Benedict et al. 2002a).  
There are, unfortunately, still ``long'' and ``short'' LMC distance
scales.
While everyone would probably agree that a distance modulus of 
\lmcdm$\,{=}\,18.35$ mag 
is consistent with the ``short scale,'' and 18.6 mag corresponds to the 
``long'' scale, the distinction between the two is somewhat artificial.
It is interesting to note that, while some techniques favor longer
or shorter distance scales on average, the measurements cover the
range with no appreciable bimodality.  Furthermore, one author may
state a distance or range of distances as being consistent with 
the long scale, while another, quoting a distance in the same
range, will state that it supports the short scale.
We regard the distinction as arbitrary; in reality, there
is a continuous range of measurements that span values significantly 
larger than the statistical uncertainties of the individual 
measurements.

There have been several compilations of LMC distances recently 
(e.g., Walker 1999; Mould et al. 2000; Benedict et al. 2002a), and
we do not review them here. 
While it is very helpful to broadly survey all the recently published
measurements, it is important to remember that a large number of 
publications of a particular value does not necessarily indicate 
correctness.  
Nor do more recent measurement necessarily deserve more trust than older 
ones.  There is some hope of resolving the debate if we look to some
recent measurements that make use of new or improved geometrical techniques
(a few are listed in Table~\ref{table2}).  
Hopefully, modest improvements in the near future will resolve the issue 
of the LMC distance, at least at the 10\% level.

The study of the SN 1987A has resulted in several 
recent geometrical distance determinations (or upper limits), ranging 
from \lmcdm$\,{<}\,18.37$ to 18.67 mag (Gould \& Uza 1997; Panagia 1999; 
Carretta et al. 2000; Benedict et al. 2002b).  The ``light
echo'' measurements are particularly important because of their
insensitivity to adopted extinction values.  
A recent spectral fitting of the expanding atmosphere of SN 1987A 
by Mitchell et al. (2002) gives a distance of $18.5\,{\pm}\,0.2$ mag.  
The SN 1987A measurements are consistent with both 
\lmcdm$\,{=}\,18.35$ and 18.50 mag.

One of the most promising techniques today makes use of detached
eclipsing binaries (DEBs), for which the orbital parameters can be 
determined and the geometrical distance derived.  
The three DEBs that have been observed in the
LMC have distances in the range of 18.30 to 18.50 mag.  
Two measurements are consistent with the short
distance scale (18.38 mag by Ribas et al. 2002; 18.30 mag by Guinan et al. 1998) 
and the other two are larger (18.50 mag by Fitzpatrick et al. 2002;
18.46 mag by Groenewegen \& Salaris 2001).  We can only conclude that
DEBs are consistent with a distance modulus to the LMC of both 18.35 and 
18.50 mag.  Measurements of many more DEBs in the LMC and other nearby galaxies
will be invaluable in helping to resolve the controversy surrounding
the distance to the LMC.

Cepheid distance measurements to the LMC generally favor the long
scale, although values as low as 18.29 mag (and as high as 18.72 mag)
have been reported (Benedict et al. 2002a).  
A good summary of the Cepheid measurements was presented by Benedict
et al. (2002a,b), who find a mean distance modulus of 18.53 mag
(average of many measurements and techniques).
This is consistent with their own measurement of $18.50\,{\pm}\,0.13$ mag 
based on new {\it HST}\ parallax measurements of $\delta$ Cephei.  
Recently, Keller \& Wood 
(2002) measured \lmcdm$\,{=}\,18.55\,{\pm}\,0.02$ mag, 
and Di~Benedetto (2002) reported
a distance modulus of $18.59\,{\pm}\,0.04$ mag.  
Although Cepheids alone cannot yet
rule out the short-scale zeropoint of 18.35 mag, the most recent measurements
are weighted toward values nearer to 18.5 mag.  
Whether or not Cepheid luminosities depend significantly 
on metallicity is an open question (see the discussion in Freedman et al.
2001).  The effect of applying the metallicity correction of 
$-0.2\,{\pm}\,0.2$
mag dex$^{-1}$  (O/H) metallicity would be to decrease the distance to the LMC 
by 0.08 mag relative to the higher-metallicity Galactic Cepheids.  
Further work with larger samples of Galactic Cepheids with higher 
metallicities than are found in the LMC is needed to resolve this issue.
One promising line of research suggests that first-overtone Cepheids
should be less sensitive to metallicity than the fundamental-mode
pulsators.  Bono et al. (2002) report overtone Cepheid distances near 
18.5 mag, in good agreement with the longer-period fundamental-mode 
Cepheids.

NGC~4258 is the only other external galaxy besides the LMC with a 
reliable geometrical distance measurement.
The orbital properties of masers around the central black hole in 
NGC~4258 can be determined to give an absolute geometrical distance 
of $7.2\,{\pm}\,0.5$ Mpc (Herrnstein et al. 1999).
Cepheids have been discovered in NGC~4258, and they provide an
independent distance measurement that is approximately 1 Mpc larger
(Maoz et al. 1999; Newman et al. 2001);  
the KP Cepheid calibration (Freedman et al. 2001) is discrepant at the
1-${\sigma}$ level if the distance to the LMC is 18.50 mag.  
The two measurements would agree if \lmcdm$=18.31$ mag.  
If the maser and Cepheid distances are both reliable, 
they could be used to rule out a distance of 18.50 to the LMC.
One alternative possibility is that the Cepheid distance
is a bit off; given the relatively small number of Cepheids detected (18),
this may not be unreasonable.  
A second possibility is that the Cepheid 
metallicity correction should have the opposite sign as that used
by the KP, as suggested by the theoretical work of Caputo, Marconi, \& 
Musella (2002).  
The recent results of Ciardullo et al. (2002) using the PNLF distance
to this galaxy suggests that the distance to the LMC should be 
reduced; both Cepheids and PNLF would be consistent with the
maser distance if \lmcdm\ were 18.3 mag.

While the original controversy between long and short distance
scales arose primarily due to differences between RR Lyrae variables
and Cepheids, the two methods are starting to converge.  
Statistical parallax measurements favor values between 18.2 and 18.3 mag, 
while Baade-Wesselink measurements prefer larger distances.  
The average value of many RR Lyrae measurements is 18.45 mag, and new data
based on {\it HST}\ parallax distances to Galactic RR Lyrae stars give 
\lmcdm\ values between 18.38 and 18.53 mag (Benedict et al. 2002a,b).
The RR Lyrae distances now agree with Cepheid distances at the 1-$\sigma$
level.

Red clump stars have been used as a distance indicator in support of
the short-distance scale (Benedict et al. 2002a).  
Red clump measurements span the range in distance modulus, from 18.07 to 
18.59 mag (Stanek, Zaritsky, \& Harris 1998; Romaniello et al. 2000).  
Recent near-IR measurements that minimize uncertainties in extinction
result in a distance modulus of $18.49\,{\pm}\,0.03$ mag (Alves et al. 2002).
Pietrzynski \& Gieren (2002) find $18.50\,{\pm}\,0.05$ mag, with a statistical
uncertainty of only 0.008 mag.
Red clump distance measurements do not yet exclude either the long 
or the short distance scales.

Many other distance measurement techniques have been applied to the LMC,
and we refer the reader to the compilation in Benedict et al. (2002a).
Most of the measurements of \lmcdm\ published this year are consistent
with a distance modulus between 18.45 and 18.55 mag, with the notable
exception of the Cepheid distance to NGC 4258, which implies 
\lmcdm$=18.31$ mag (although uncertainties in the metallicity correction
to Cepheid distances lessen the significance of the discrepancy,
as explained by Caputo et al. 2002).  
Since the NGC 4258 Cepheid distance is only discrepant 
at the 1-$\sigma$ level, we believe that a change from the LMC distance 
of 18.50 mag is not justified at the present time.  

Resolving the debate between the long
and short distances to the LMC will not necessarily reduce the uncertainty
in the Hubble constant.  The differences between techniques have usually
exceeded the quoted uncertainties, both systematic and statistical.
The systematic uncertainty in \lmcdm\ adopted by the KP team was 0.13
mag.  Even if we choose the best results from the different techniques
that fall closest to the adopted modulus of 18.50 mag, the scatter will 
likely still be ${>}\,0.13$ mag.  Furthermore, it is likely that any 
individual measurement, taken on its own as the most reliable available, 
will have a total uncertainty no better than 0.1 mag.  Reducing the 
uncertainty in the LMC distance modulus will require more than
the elimination of systematic errors that are not completely accounted
for in the current set of uncertainty estimates.  It is, however, a
crucial step in our progress toward improving the precision of the
distance ladder techniques.

% Table 2:  LMC distances
\begin{table}
\caption{A Summary of Recently Published LMC Distances}
%\tiny
%\footnotesize
\scriptsize
\begin{tabular}{lll}
\hline \hline
{Technique} & {\lmcdm} (mag)& {Reference} \\
\hline
Cepheids/masers (NGC 4258)\dotfill& $18.31\pm0.11$ & Newman et al. 2002 \\
Cepheids ($\delta$ Cep)\dotfill & $18.50\pm0.13$  & Benedict et al. 2002b \\
Cepheids (mean of many techniques)\dotfill & $18.53$ & Benedict et al. 2002a,b\\
Eclipsing binaries\dotfill      & $18.38\pm0.08$  & Ribas et al. 2002\\
                                & $18.50\pm0.05$  & Fitzpatrick et al. 2002\\
                                & $18.46\pm0.07$  & Groenewegen \& Salaris 2001\\
                                & $18.30\pm0.07$  & Guinan et al. 1998\\ 
SN 1987A\dotfill                & $18.5\pm0.2$   & Mitchell et al. 2002\\
RR Lyr\dotfill                  & 18.38 to 18.53  & Benedict et al. 2002a\\
RR Lyr (mean of many techniques)\dotfill& $18.45\pm0.08$  & Benedict et al. 2002a,b\\
Red clump\dotfill               & $18.49\pm0.03$ & Alves et al. 2002\\
Red clump\dotfill               & $18.50\pm0.008\pm0.05$ & Pietrzynski \& Gieren 2002\\
\hline \hline
\end{tabular}
\label{table2}
\end{table}

\subsection{Metallicity Corrections to Cepheid Luminosities}

The possibility that the luminosity of a Cepheid variable star depends on
its metallicity is one of the most significant remaining uncertainties.
Most of the distant galaxies with known Cepheid distances have metallicities
similar to those of the Galactic Cepheids, and significantly higher 
metallicity than
the LMC Cepheids.  The magnitude of the metallicity correction is not very 
important, provided that it is applied consistently.  
At the present time it is not clear if a metallicity
correction is justified or not (Udalski et al. 2001; Caputo et al. 2002; 
Jensen et al. 2003).

Most of the Cepheid distance measurements published, including 
the KP papers prior to Freedman et al. (2001), have no
metallicity correction applied.  
For the final KP results, Freedman et al. adopted a metallicity correction
of $-0.2$ mag dex$^{-1}$  in (O/H) (Kennicutt et al. 1998).
Since both the old and new Cepheid calibrations
use the same LMC distance \lmcdm$\,{=}\,18.50$ mag, 
it should be noted that a direct
comparison between the old calibration (Ferrarese et al. 2000b) 
and the Freedman et al. (2001) calibration, which includes 
the metallicity correction, requires an offset
of ${\sim}0.08$ mag due to the difference between Galactic and LMC Cepheid
metallicities.  In other words, the difference between the Mould et al. (2000)
and Freedman et al. (2001) values of \Ho\ would be 4\% larger 
than reported if Galactic and LMC Cepheids were metallicity-corrected
the same as Cepheids in the more distant galaxies.

The Freedman et al. (2001) distances derived using the new PL 
calibration can be used without the metallicity correction,
with a corresponding increase in \Ho.
Because of the distance-dependent nature of the change to the new
PL relation, the Hubble constant that results from using no
metallicity correction depends on which Cepheid calibrators are
used to tie to distant secondary techniques.
For the case of $I$-band SBFs, the new PL relation without
metallicity corrections results in an increase in the Hubble constant of
8\% over the original KP calibration, and 5\% over the new PL relation
with metallicity corrections included.

\subsection{Systematic Photometric Uncertainties}

One significant source of systematic error in the data taken with 
WFPC2 has been the uncertainty in the photometric zeropoint,
which the KP team estimate at 0.09 mag.
Extinction corrections also contribute to the photometric
uncertainties in Cepheid measurements.  More details can be found
in the KP papers (see Freedman et al. 2001).

Blending of Cepheids with other stars is another potential source of
systematic uncertainty.  
If blending is significant, the Cepheid
distances to the most distant galaxies surveyed may be underestimated
by 10 to 20\% (Mochejska et al. 2000).
Gibson, Maloney, \& Sakai (2000) found no evidence of a trend in
residuals with distance and no difference between the WF and PC
camera measurements, which have different spatial sampling and therefore
different sensitivities to blending.
Ferrarese et al. (2000c) also found no significant offset in the Cepheid
photometry due to crowding.  They based their uncertainty estimate of 
0.02 mag on tests in which they added artificial stars to
their images and processed them in the same way as the real stars.
The strict criteria for selecting and measuring Cepheids used by the KP
team can explain the small effect of blending on the photometry of 
the artificial stars added.

Although improved spatial resolution helps minimize the effects of blending,
it is not complete protection.  Physical companions to Cepheids cannot
be resolved.
Furthermore, the effect of blending on Cepheid distances is not limited to a
single companion.  The surface brightness fluctuations in the 
underlying population are a background with structure on the scale of
the point-spread function that can make the Cepheid
look brighter or fainter.  In the most distant galaxies, there will be
a slight detection bias in favor of Cepheids superimposed on bright
fluctuations.  The fluctuations are very red, so
not only will the brightness of the Cepheid be overestimated, but the
color observed will be too red.  The corresponding extinction correction 
will be larger than it should be, enhancing the overestimate of the Cepheid
luminosity and underestimate of the distance.

\section{Peculiar Velocities}

Any measurement of the Hubble constant
relies on both distance {\it and} velocity measurements.
Within 50 Mpc, the clumpy distribution of mass leads to peculiar 
velocities that can be larger than the Hubble expansion velocity.  
It is therefore critical that recession velocities
within 50 Mpc be corrected depending on where the galaxy lies
relative to the Virgo cluster, Great Attractor, and so forth.
Differences in how these corrections are applied has led to 
significant differences in measured values of \Ho.  Half the
difference between the KP SBF Hubble constant and that of Tonry
et al. (2000) is due to differences in the velocity model.
Furthermore, there is some evidence suggesting that 
the local Hubble expansion rate is slightly larger than the global
value, which would be the natural result of the local Universe
being slightly less dense than the global average (Zehavi et al. 1998;
Jensen et al. 2001).  While the evidence is far from conclusive
(Giovanelli, Dayle, \& Haynes 1999; Lahav 2000), 
it reinforces the importance of measuring
the expansion rate of the Universe as far out as possible.
There is obviously great advantage in measuring \Ho\
at distances large enough to be free of peculiar velocities.  
Beyond 100 Mpc, even the largest peculiar velocities (${\sim}1500$ \kms) 
are only a fraction of the Hubble velocity (${\sim}7000$ \kms).  
The Ia SNe, SBF, Tully-Fisher, and FP Hubble constant
measurements made beyond 100 Mpc presented in the previous sections all 
show the excellent consistency 
expected from a solid tie to the distant Hubble flow.

\section{Bypassing the Distance Ladder \label{bypass}}

Two techniques, gravitational lens time delays and the Sunyaev-Zel'dovich (SZ)
effect, promise to provide  measurements of \Ho\
at significant redshifts independent of the calibration of the local 
distance scale.  Both of these techniques are discussed in detail 
elsewhere in this volume (Kochanek \& Schechter and Reese, this volume).

Time delay measurements in multiple-image gravitational 
lens systems can provide a geometrical distance and Hubble constant
provided the mass distribution is known in the radial region
between the images of the gravitationally lensed quasar
(Kochanek 2002, 2003).
Recent time-delay measurements give \Ho$\,{\approx}\,60$ \kmsmpc, 
and are somewhat lower than \Ho\ found using Cepheid-calibrated 
secondary distance indicators.
Treu \& Koopmans (2002) reported \Ho$\,{=}\,59\,^{+12}_{-7}\,{\pm}\,3$ 
\kmsmpc\ for an $\Omega_{\Lambda}\,{=}\,0.7, \Omega_{m}\,{=}\,0.3$ Universe.  
Fassnacht et al. (2002) found values between 61 and 65 for the same
cosmological model.  Cardone et al. (2002) found 
\Ho$\,{=}\,58\,^{+17}_{-15}$ \kmsmpc, in good agreement with the others.
Kochanek (2002) showed
that values of \Ho\ between 51 and 73 \kmsmpc\ are possible;
the lower limit corresponds to cold dark matter $M/L$ concentrations,
while the upper limit is set by the constant-$M/L$ limit.  
Now that more accurate time delays have been measured for ${\sim}5$ 
systems using radio, optical, and X-ray observations, 
the distribution of mass in the 
lensing galaxy is the largest remaining uncertainty in gravitational
lens measurements of the Hubble constant.

The SZ effect at submillimeter wavelengths, 
when combined with X-ray measurements of the hot gas in galaxy clusters, 
can be exploited to determine the angular diameter distance to the cluster
(Carlstrom, Holder, \& Reese 2002; Reese et al. 2002).  
To date, there are 38 SZ distance measurements extending to redshifts of
$z\,{=}\,0.8$.  A fit to the 38 measurements yields \Ho$\,{=}\,60\,{\pm}\,3$ 
\kmsmpc, with an additional systematic uncertainty of order 30\%
(Carlstrom et al. 2002).  The primary systematic uncertainties arise
from cluster structure (clumpiness and departures from isothermality)
or from point-source contamination.  The SZ Hubble constant is also
a function of the mass and dark energy density of the Universe;
the value presented here assumes $\Omega_{m}\,{=}\,0.3$ and 
$\Omega_{\Lambda}\,{=}\,0.7$ (Reese et al. 2002).
Future SZ surveys are expected to discover hundreds of new clusters,
which should greatly improve our understanding of the density and
expansion rate of the Universe to $z{\approx}2$ (Carlstrom et al. 2002).

Using models to predict the absolute luminosity of a standard candle 
is another way to sidestep the issues with empirical distance scale 
calibrations.
Both Type Ia and II SNe, along with SBFs, can be primary distance indicators.
We usually choose to calibrate SNe and SBFs empirically using Cepheids
because of the acknowledged uncertainties in the many model parameters.
The models are good enough, however, to provide some constraints on
the distance scale and the quality of the Cepheid calibration.

Recent models of Type Ia SNe are now detailed enough to predict the
absolute luminosity of the burst and therefore allow distances to be
derived directly.  The results reported by H\"oflich \& Khokhlov (1996),
for example, show that Ia SN models are consistent with a 
Hubble constant of $67\,{\pm}\,9$
\kmsmpc.  While there are clearly many details in the explosion 
models that must be carefully checked against observations, it is
reassuring that the predictions are in the right ball park.
Type II SNe  have also been used as primary standard candles
using a theoretically calibrated expanding-photosphere 
technique (Schmidt et al. 1994; Hamuy 2001).
Schmidt et al. found \Ho$\,{=}\,73\,{\pm}\,6\,{\pm}\,7$ \kmsmpc, 
independent of the Cepheid calibration.  
Hamuy's (2001) updated measurement using the
same technique yielded $67\,{\pm}\,7$ \kmsmpc.

SBFs are proportional to the second moment of the stellar luminosity
function.  Standard stellar population models can be integrated to 
predict SBF magnitudes for populations with particular ages and 
metallicities, and then compared to observations 
(Blakeslee, Vazdekis, \& Ajhar 2001; Liu, Charlot, \& Graham 2000; 
Liu, Graham, \& Charlot 2002).  
In the $I$ band, SBF comparisons with stellar population models 
would agree with observations better if the original KP 
Cepheid zeropoint were fainter by $0.2\,{\pm}\,0.1$ mag 
(Blakeslee 2002), which would make \Ho\ 10\% larger.
Jensen et al. (2003) found that $H$-band SBFs were entirely 
self-consistent with both the Vazdekis (1999, 2001) 
and Bruzual \& Charlot (1993) models 
when the calibration of Freedman et al. (2001) was used without 
metallicity corrections.
The Hubble constant implied by the IR population
models is 8\% higher than that determined using the original KP
Cepheid calibration of Ferrarese et al. (2000a) and 5\% higher than
that of Freedman et al. (2001) with the metallicity correction.

\section{Probability Distributions and Systematic Uncertainties}

As distance measurement techniques mature, two things generally happen:
first, the number of measurements increases, reducing the statistical
uncertainty, and second, the larger data sets make it possible to 
correct for secondary effects that modify the brightness of the
standard candle.  The result is a reduction in the statistical
uncertainty to the point that systematic effects start to dominate.
This is certainly true for Cepheids and the secondary distance
indicators calibrated using them.  As shown in previous sections,
systematic uncertainties in the Cepheid distances dominate statistical
uncertainties and the systematic differences between different
secondary techniques.  
Addressing systematic uncertainties is a difficult job
that cannot proceed until a sufficiently large number of  measurements has 
been made to understand the intrinsic
dispersion in the properties of a standard candle.

Some of the techniques discussed have not yet been applied to enough
systems to conclusively say that small number statistics are not
an issue, even though the formal statistical uncertainty of an
individual measurement might be small.  Gravitational lens time delays
have only been measured in five systems, for example.  SZ
measurements are only now reaching large enough samples to start
addressing the systematic uncertainties.  Fortunately, the sample sizes
will increase significantly as surveys to find and measure more lensed
quasars and SZ clusters proceed.  Supernovae of all types
are rare enough that finding enough nearby SNe for calibration
purposes has required using limited and often old, unreliable data.

The probability distributions of systematic uncertainties is not always
known, and is frequently not Gaussian.  For example, the range of 
\Ho\ values permitted by the gravitational lens measurements is 
set by systematic uncertainties in the lens mass distribution.
The extremes are rather rigidly limited by constraints on the possible
fraction and distribution of dark matter in galaxies.
The probability distribution is therefore rather close to a top hat,
with ``sigma limits'' that are not at all Gaussian 
(e.g., 2-${\sigma} < 2{\times}\sigma$).  
It is clearly not appropriate to add errors of this nature in quadrature.  

Many researchers have maintained separate accounting of systematic and
random uncertainties when possible (cf., Ferrarese et al. 2000a).  
Even this approach
requires the addition of different systematic uncertainties in quadrature,
assuming that individual systematic uncertainties are independent and
Gaussian in nature.
In many cases this is probably justified; in others, simply reporting
a range of possible values is more appropriate.  The problem with 
reporting such ranges as a systematic uncertainty is that they are often
viewed as being overly pessimistic by the casual reader who regards
them as ``1-{$\sigma$}'' uncertainties.  Given the difficulty in
comparing very different systematic uncertainties, it is 
probably premature to judge the SZ and gravitational lens results as 
being inconsistent with the 
Cepheid-based distance indicators at the present time when the 
number of measurements is still rather few and the systematics have
not been explored in great detail.

\section{Future Prospects}

The secondary distance indicators, mostly calibrated using Cepheids, 
are all in good agreement when calibrated uniformly.  These imply a 
Hubble constant between 70 and 75 \kmsmpc.  Many other techniques
that do not rely on the calibration of the traditional distance ladder
generally agree with this result at the 1-${\sigma}$ level; further
concordance between independent techniques will require a careful analysis
of systematic uncertainties and new survey data that will become available
in the next few years.
Improvements in the Cepheid calibration will require a better zeropoint 
and larger samples covering a range of period and metallicity. 
Near-IR photometry will help reduce uncertainties due to extinction.
High-resolution imaging will help reduce blending and allow 
measurements of Cepheids in more distant galaxies.  Improved
photometry and excellent resolution will make the ACS and WFPC3 
on {\it HST}\ powerful Cepheid-measuring instruments.

{\it SIM}\ and {\it GAIA}\ are two astrometry satellite missions planned by NASA and
ESA that will help reduce systematic uncertainties in the extragalactic
distance scale by providing accurate (1\%) parallax distances for a
significant number of Galactic Cepheids.
{\it SIM}\ and {\it GAIA}\ will allow us to calibrate the 
Cepheid zeropoint, PL relation, and metallicity corrections without 
having to rely on the LMC sample or the distance to the LMC.  
We are optimistic that the debate over the ``long'' and ``short''
distance scales for the LMC will soon be behind us.  While this will
be a significant milestone, it will not help much to reduce the statistical
uncertainty in the measured value of the Hubble constant.  It will,
however, remove one persistent source of systematic error.  
To achieve a Hubble constant good to 5\% using a distance estimator 
tied to the LMC will require much more accurate geometrical 
measurements, and a reduction of the other systematic uncertainties as well.

With {\it SIM}\ and {\it GAIA}, the calibration of several variable star distance scales
in addition to the Cepheids will be solidified.  These include RR Lyrae
variables,
delta Scuti stars, and shorter-period overtone pulsators.  The 
increased sensitivity and spatial resolution of the next generation of
large ground and space telescopes will
allow us to detect these fainter variables in the distant galaxies
in which only Cepheids are currently detectable.  Other variable stars
will allow us to resolve questions about bias that arise when only the 
very brightest members of the Cepheid population are detected and used 
to determine the distance.

The number of ways to bypass the Cepheid rung of the distance ladder
will increase dramatically when
{\it SIM}\ and {\it GAIA}\ allow us to calibrate a number of secondary distance indicators
directly from statistical parallax distance measurements to M31, M32, and M33.
Techniques like SBF, Tully-Fisher, FP, GCLF, PNLF, 
and so forth, have already been used to determine accurate relative distances
between the Local Group members M31, M32, and M33 and more distant galaxies.  
By determining their distances directly from statistical parallax 
measurements, 
the systematic uncertainties in the Cepheid calibration and LMC distance 
will be avoided altogether.

Improvements in techniques that bypass the local distance ladder
and secondary distance indicators are imminent.  Larger samples
of well-measured gravitational lens time delays and SZ
clusters will help reduce statistical uncertainties and provide insight
into the systematics.
Better mass models for gravitational lenses will not only lead to better
determinations of \Ho, but also be valuable in constraining the quantity
and distribution of dark matter in galaxies.
The increasing number of SZ measurements will lead to a better 
understanding of galaxy cluster structure and evolution.

The number of direct geometrical distance measurements to nearby galaxies
will also increase.  The masers detected in NGC~4258 must also exist in
other galaxies.  A larger sample of detached eclipsing binaries, both in
the LMC and in other galaxies, will help overcome the systematic 
uncertainties and provide more consistent distances.  These techniques,
like SNe in nearby galaxies, are limited by small-number statistics.
An individual measurement may seem reliable, but until more are found,
our confidence in them will be limited.

Several new synoptic and survey facilities are currently being planned
that will discover many thousands of SNe.  The {\it SNAP}\ satellite
will discover thousands of SNe over its lifetime.  It will be able to 
measure optical and near-IR brightnesses and collect spectra for SN
classification.  The Large Synoptic Survey Telescope (LSST) and PanSTARRs
survey telescopes, currently in the planning stages, will discover 
hundreds of thousands of Ia SNe every year.  The synoptic telescopes
will also reveal a multitude of faint variable stars in the Galaxy.  
With this wealth of data,
systematic uncertainties can be addressed and the expansion rate of 
the Universe determined as a function of redshift to $z\,{>}\,1$.  
We will only be limited by our ability to follow up the SN discoveries
to determine reliable distances.  

Perhaps the best determination of \Ho\ in the future will come from
the combination of multiple joint constraints, just as the conclusions 
regarding the $\Omega_{\Lambda}=0.7, \Omega_{m}=0.3$ Universe came from 
merging the Type Ia SN results with the measurements of 
$\Omega_{tot}\,{=}\,1.0$ from the cosmic microwave background experiments.  
For example, both $H_{0}t_{0}$ and $\Omega_{b}h^2$ are now known to
5\%.  Other examples of joint
constraints that include \Ho\ are described elsewhere in this volume.

\vspace{0.3cm}
{\bf Acknowledgements}.
J. Jensen acknowledges the support of the Gemini Observatory,
which is operated by the Association of Universities for Research in
Astronomy, Inc., on behalf of the international Gemini partnership of 
Argentina, Australia, Brazil, Canada, Chile, the United Kingdom, and the 
United States of America.


\begin{thereferences}

\bibitem{}
Ajhar, E. A., Tonry, J. L., Blakeslee, J. P., Riess, A. G., \&
Schmidt, B. P. 2001, \apj, 559, 584

\bibitem{}
Alves, D. R., Rejkuba, M., Minniti, D., \& Cook, K. H. 2002, \apj,
573, L51

\bibitem{}
Benedict, G. F., et al. 2002a, \aj, 123, 473

\bibitem{}
------. 2002b, \aj, 124, 1695

\bibitem{}
Blakeslee, J. P. 2002, in A New Era In Cosmology, 
ed. T. Shanks \& N. Metcalfe (Chelsea: Sheridan Books)

\bibitem{}
Blakeslee, J. P., Ajhar, E. A., \& Tonry, J. L. 1999, in 
Post-Hipparcos Cosmic Candles, ed. A. Heck \& F. Caputo 
(Dordrecht: Kluwer), 181

\bibitem{}
Blakeslee, J. P., Lucey, J. R., Tonry, J. L., Hudson, M. J.,
Narayanan, V. K., \& Barris, B. J. 2002, \mnras, 330, 443

\bibitem{}
Blakeslee, J. P., Vazdekis, A., \& Ajhar, E. A. 2001, \mnras, 320, 193

\bibitem{}
Bono, G., Groenewegen, M. A. T., Marconi, M., \& Caputo, F. 2002, \apj, 574, L33

\bibitem{}
Bruzual A., G., \& Charlot, S. 1993, \apj, 405, 538

\bibitem{}
Caputo, F., Marconi, M., \& Musella, I. 2002, \apj, 566, 833 

\bibitem{}
Cardone, V. F., Capozziello, S., Re, V., \& Piedipalumbo, E. 2002,
\aap, 382, 792

\bibitem{}
Carlstrom, J. E., Holder, G. P., \& Reese, E. D. 2002, \araa, 40, 643

\bibitem{}
Carretta, E., Gratton, R. G., Clementini, G., \& Pecci, F. F. 2000,
\apj, 533, 215

\bibitem{}
Ciardullo, R., Feldmeier, J. J., Jacoby, G. H., De Naray, R. K., 
Laychak, M. B., \& Durrell, P. R. 2002, \apj, 577, 31

\bibitem{}
Di Benedetto, G. P. 2002, \aj, 124, 1213

%\bibitem{}
%Faber, S. M., Wegner, G., Burstein, D., Davies, R. L., Dressler, A., 
%Lynden-Bell, D., \& Terlevich, R. J. 1989, \apj, 69, 763

\bibitem{}
Fassnacht, C. D., Xanthopoulos, E., Koopmans, L. V. E., \& Rusin, D.
2002, \apj, 581, 823

\bibitem{}
Ferrarese, L., et al. 2000a, \apj, 529, 745 

\bibitem{}
------. 2000b, \apjs, 128, 431

\bibitem{}
Ferrarese, L., Silbermann, N. A., Mould, J. R., Stetson, P. B., 
Saha, A., Freedman, W. L., \& Kennicutt, R. C., Jr. 2000c, \pasp, 112, 117

\bibitem{}
Fitzpatrick, E. L., Ribas, I., Guinan, E. F., DeWarf, L. E., Maloney, F. P., 
\& Massa, D. 2002, \apj, 564, 260

\bibitem{}
Freedman, W.~L., et al. 2001, \apj, 553, 47 

\bibitem{}
Freedman, W.~L., \& Madore, B.~F. 1990, \apj, 365, 186

%\bibitem{}
%Gibson, B. K., et al. 1999, \apj, 512, 48

\bibitem{}
Gibson, B. K., et al. 2000, \apj, 529, 723

\bibitem{}
Gibson, B. K., Maloney, P. R., \& Sakai, S. 2000, \apj, 530, L5

\bibitem{}
Gibson, B. K., \& Stetson, P. B. 2001, \apj, 547, L103 

\bibitem{}
Giovanelli, R., Dale, D. A., \& Haynes, M. P. 1999, \apj, 525, 25

\bibitem{}
Giovanelli, R., Haynes, M. P., Herter, T., Bogt, N. P., Da Costa, L. N., 
Freudling, W., Salzer, J. J., \& Wegner, G. 1997, \aj, 113, 22

\bibitem{}
Gould, A., \& Uza, O. 1997, \apj, 494, 118

\bibitem{}
Groenewegen, M. A. T., \& Salaris, M. 2001, \aap, 366, 752

\bibitem{}
Guinan, E. F., et al.  1998, \apj, 509, L21

\bibitem{}
Hamuy, M. 2001, Ph.D. Thesis, Univ. Arizona

\bibitem{}
Hamuy, M., Phillips, M. M., Maza, J., Suntzeff, N. B., Schommer, R. A.,
\& Aviles, R. 1995, \aj, 109, 1669

\bibitem{}
Hamuy, M., Phillips, M. M., Suntzeff, N. B., Schommer, R. A., Maza, J., \& Aviles, R.
1996, \aj, 112, 2398

\bibitem{}
Herrnstein, J. R., et al.  1999, Nature, 400, 539

\bibitem{}
H\"oflich, P., \& Khokhlov, A. 1996, \apj, 457, 500

\bibitem{}
Hudson, M. J., Lucey, J. R., Smith, J. R., Schlegel, D. J., 
\& Davies, R. L. 2001, \mnras, 327, 265

\bibitem{}
Jensen, J. B., Tonry, J. L., Barris, B. J., Thompson, R. I., 
Liu, M. C., Rieke, M. J., Ajhar, E. A., \& Blakeslee, J. P.  2003, 
\apj, 583, 712

\bibitem{}
Jensen, J. B., Tonry, J. L., Thompson, R. I., Ajhar, E. A., Lauer, T. R., 
Rieke, M. J., Postman, M., \& Liu, M. C. 2001, \apj, 550, 503 

\bibitem{}
Keller, S. C., \& Wood, P. R. 2002, \apj, 578,144

\bibitem{}
Kelson, D.~D., et al. 2000, \apj, 529, 768

\bibitem{}
Kennicutt, R. C., Jr., et al. 1998, \apj, 498, 181 

\bibitem{}
Kim, A. G., et al. 1997, \apj, 476, L63

\bibitem{}
Kochanek, C. S. 2002, \apj, 578, 25

\bibitem{}
------. 2003, \apj, submitted

\bibitem{}
Kundu, A., \& Whitmore, B. C. 2001, \aj, 121, 2950

\bibitem{}
Lahav, O. 2000, in Cosmic Flows 1999: Towards an Understanding of Large-Scale 
Structure, ed. S. Courteau, M. A. Strauss, \& J. A. Willick (Chelsea: 
Sheridan Books), 377

\bibitem{}
Lauer, T. R., Tonry, J. L., Postman, M., Ajhar, E. A., \& Holtzman, J. A. 
1998, \apj, 499, 577

\bibitem{}
Liu, M. C., Charlot, S., \& Graham, J. R. 2000, \apj, 543, 644

\bibitem{}
Liu, M. C., \& Graham, J. R. 2001, \apj, 557, L31

\bibitem{}
Liu, M. C., Graham, J. R., \& Charlot, S. 2002, \apj, 564, 216

\bibitem{}
Macri, L. M., et al. 2001, \apj, 549, 721

\bibitem{}
Maoz, E., Newman, J. A., Ferrarese, L., Stetson, P. B., Zepf, S. E., 
Davis, M., Freedman, W. L., \& Madore, B. F. 1999, Nature, 401, 351

\bibitem{}
Mitchell, R. C., Baron, E., Branch, D., Hauschildt, P. H., Nugent, P. E.,
Lundqvist, P., Blinnikov, S., \& Pun, C. S. J. 2002, \apj, 574, 293

\bibitem{}
Mochejska, B. J., Macri, L. M., Sasselov, D. D., \& Stanek, K. Z. 2000,
\aj, 120, 810

\bibitem{}
Mould, J. R., et al. 2000, \apj, 529, 786

\bibitem{}
Newman, J. A., Ferrarese, L., Stetson, P. B., Maoz, E., Zepf, S. E., 
Davis, M., Freedman, W. L., \& Madore, B. F., 2001, \apj, 553, 562

\bibitem{}
Okon, V. M. M., \& Harris, W. E. 2002, \apj, 567, 294

\bibitem{}
Paczy\'nski, B. 2001, Acta Astron., 51, 81

\bibitem{}
Panagia, N. 1999, in IAU Symp. 190, New Views of the Magellanic Clouds,
ed. Y.-H. Chu et al. (Dordrecht: Kluwer), 549

\bibitem{}
Parodi, B. R., Saha, A., Sandage, G. A., \& Tammann, G. A. 2000,
\apj, 540, 634

\bibitem{}
Perlmutter, S. et al. 1997, \apj, 483, 565

\bibitem{}
------. 1998, Nature, 391, 51

\bibitem{}
Phillips, M. M. 1993, \apj, 413, L105

\bibitem{}
Pietrzynski, G., \& Gieren, W. 2002, \aj, 124, 2633

\bibitem{}
Reese, E. D., Carlstrom, J. E., Joy, M., Mohr, J. J., Grego, L.,
\& Holzapfel, W. L. 2002, \apj, 581, 53

\bibitem{}
Ribas, I., Fitzpatrick, E. L., Maloney, F. P., \& Guinan, E. F. 2002,
\apj, 574, 771

\bibitem{}
Riess, A. G., Press, W. H., \& Kirshner, R. P. 1996, \apj, 473, 88

\bibitem{}
Romaniello, M., Salaris, M., Cassisi, S., \& Panagia, N. 2000, \apj, 530, 738 

\bibitem{}
Saha, A., Sandage, A., Tammann, G. A., Dolphin, A. E., Christensen, J., 
Panagia, N., \& Macchetto, F. D. 2001, \apj, 562, 314

\bibitem{}
Sakai, S., et al. 2000, \apj, 529, 698

\bibitem{}
Sebo, K. M., Rawson, D., Mould, J., Madore, B. F., Putman, M. E., 
Graham, J. A., Freedman, W. L., Gibson, B. K., \& Germany, L. M. 2002,
\apjs, 142, 71

\bibitem{}
Schmidt, B. P., et al. 1998, \apj, 507, 46

\bibitem{}
Schmidt, B. P., Kirshner, R. P., Eastman, R. G., Phillips, M. M.,
Suntzeff, N. B., Hamuy, M., Maza, J., \& Aviles, R. 1994, \apj, 432, 42

\bibitem{}
Stanek, K. Z., Zaritsky, D., \& Harris, J. 1998, \apj, 500, L141

\bibitem{}
Tanvir, N. R., Ferguson, H. C., \& Shanks, T. 1999, \mnras, 310, 175

\bibitem{}
Tonry, J. L., Blakeslee, J. P., Ajhar, E. A., \& Dressler, A. 2000,
\apj, 530, 625

\bibitem{}
Tonry, J. L., Dressler, A., Blakeslee, J. P., Ajhar, E. A., Fletcher, A. B., 
Luppino, G. A., Metzger, M. R., \& Moore, C. B. 2001, \apj, 546, 681 

\bibitem{}
Tonry, J. L., \& Schneider 1988, \aj, 96, 807

\bibitem{}
Treu, T., \& Koopmans, V. E. 2002, \mnras, 337, L6

\bibitem{}
Udalski, A., Soszynski, I., Szymanski, M., Kubiak, M., 
Pietrzynski, G., Wozniak, P., \& Zebrun, K. 1999a, Acta Astron., 49, 223

\bibitem{}
Udalski, A., Szymanski, M., Kubiak, M., Pietrzynski, G., 
Soszynski, I., Wozniak, P., \& Zebrun, K. 1999b, Acta Astron., 49, 201

\bibitem{}
Udalski, A., Wyrzykowski, L., Pietrzynski, G., Szewczyk, O.,
Szymanski, M., Kubiak, M., Soszynski, I., \& Zebrun, K. 2001,
Acta Astron., 51, 221

\bibitem{}
Vazdekis, A. 1999, \apj, 513, 224

\bibitem{}
------. 2001, Ap\&SS, 276, 921

\bibitem{}
Walker, A. 1999, in Post-Hipparcos Cosmic Candles, ed. A. Heck \& F. Caputo
(Dordrecht: Kluwer), 125

\bibitem{}
Willick, J. A., \& Batra, P. 2001, \apj, 548, 564

\bibitem{}
Zehavi, I., Riess, A. G., Kirshner, R. P., \& Dekel, A. 1998, \apj, 503, 483

\end{thereferences}
\end{document}